# Flexible delivery of broadband, 100-fs mid-infrared pulses in the water-absorption band using hollow-core photonic crystal fibre

W. LIN,[1,2,3,†] Z. LI,[1,3,†] Y. TENG,[1,2,3] J. HUANG,[1,3,5,*] Y. ZHAO,[1,3] Z. LUO,[1,3] W. SUN,[1,3] C. JIANG,[1,3] R. YING,[1] Y. ZHENG,[4] X. JIANG,[1,3,4,6,*] AND M. PANG[1,2,3,7,*]

[1]*Russell Centre for Advanced Lightwave Science, Shanghai Institute of Optics and Fine Mechanics and Hangzhou Institute of Optics and Fine Mechanics, Hangzhou 311400, China*
[2]*School of Physics and Optoelectronic Engineering, Hangzhou Institute for Advanced Study, University of Chinese Academy of Sciences (UCAS), Hangzhou 310024, China.*
[3]*Shanghai Institute of Optics and Fine Mechanics (SIOM), Chinese Academy of Sciences (CAS), Shanghai 201800, China*
[4]*iFiber (Ningbo) Optoelectronics Technology Co., LTD., Ningbo 315000, China*
[†]*These authors contributed equally.*
[5]*jiapenghuang@siom.ac.cn*
[6]*xin.jiang@r-cals.com*
[7]*pangmeng@siom.ac.cn*

**Abstract:** High quality free-space and over-fibre transmission of mid-IR light is limited by factors such as material-related absorption, diffraction, light leakage and nonlinearity. Conventional vacuum apparatus can be utilized for high-quality laser-beam delivery to address these issues, the deployment of such apparatus would, however, increase the system complexity, being detrimental to their practical applications. Here we report the successful use of evacuated hollow-core photonic crystal fibre (PCF) to flexibly transmit ultrafast mid-IR pulses over several meters, while preserving exceptional spatial, spectral and temporal fidelity. The PCF was engineered to feature a low-loss transmission band within the water absorption range, and an evacuated 5-m length was used to transmit Watt-level, 100 fs pulses centred at ~2.8 μm. A comparison between free-space transmission and air-filled PCF highlights the superior performance of the evacuated hollow-core PCF, indicating its strong suitability for the flexible delivery of sub-ps laser pulses in the mid-IR.

## 1. Introduction

High-power mid-infrared (mid-IR) light sources with sub-ps pulse durations and broad spectral coverage have important applications for example in advanced spectroscopy [1-4], material processing [5, 6], surgery [7], and remote sensing [8]. High-fidelity in-fibre transmission of mid-IR pulses extends the flexibility of systems used in remote sensing [9], industrial processing and biomedical diagnostics. Fibre delivery of light for mid-IR spectroscopy is particularly attractive in harsh or hazardous environments [10], where the operation and maintenance of advanced lasers is difficult. In material processing and surgery, fibre delivery is poised to replace bulky beam-delivery arms [11-15], thereby improving the flexibility and reducing the footprint of such systems.

While mid-IR optical fibres made from fluoride, tellurite and chalcogenide glasses, as well as those featuring polycrystalline silicon cores, have been extensively investigated [16-18], achieving high-fidelity transmission of Watt-level sub-ps mid-IR laser pulses over several metres have proven elusive. Two primary obstacles are present. First, in the mid-IR region air has many molecular absorption lines [3, 19], which results in power loss and degradation of both temporal and spatial beam profiles during free-space propagation [19]. Second, high peak power in solid-core fibres made from mid-IR glass leads to nonlinear effects that cause catastrophic spectral and temporal distortion to the pulse profile [20]. Although hollow core photonic crystal fibres (PCF) made from mid-IR glass have been developed [21-23], fabrication difficulties have hindered their practical application. In recent years, hollow-core PCF made from silica glass has been successfully used for low-loss transmission of mid-IR light out to 7 μm [24]. Compared to mid-IR fibres with solid glass cores, hollow core PCF has a significantly higher optical damage threshold and enhanced mechanical strength. Notably, the maturation of the stack-and-draw method enables the fabrication of high-quality fibres in kilometer lengths [13, 25-28]. To date, a range of silica hollow-core PCFs have been reported for the mid-IR [14, 29, 30], exhibiting broad spectral coverage and low loss: 0.05 dB/m at 3.6 μm [30], 0.04 dB/m at 4 μm [29] and 0.24 dB/m at 4.6 μm [14]. However, achieving distortion-free fibre delivery of high-energy sub-ps pulses at mid-IR wavelengths has remained a challenge.

In this article, we report what we believe to be the first successful fibre delivery of ~100 fs pulses in the water-window (~2.8 μm) with peak power ~300 kW, while maintaining high spectral, temporal and spatial fidelity. Absorption and nonlinear effects are minimised by using an evacuated hollow-core PCF, resulting in >70% overall transmission over 5 m, along with excellent preservation of the spectral and temporal pulse shape after pre-compensating for fibre dispersion.

## 2. Hollow-core PCF with broad transmission window at mid-IR

The characteristics of the hollow-core PCF are summarised in Fig. 1. Fabricated using the stack-and-draw technique, it consists of a central hollow core of diameter ~103 μm, surrounded by a single ring of eight capillaries with inner diameter ~40 μm and wall thickness ~805 nm. As shown in Fig. 1(b), the lowest order anti-crossing between the core mode and resonances in the capillary walls occurs at ~1600 nm, in good agreement with theoretical predictions [12]. At wavelengths longer than 1600 nm the transmission window remains uninterrupted, reaching 4.5 μm. Light from a mid-IR supercontinuum source (ELECTRO MIR4.8 from Leukos) was launched into a 50 m length of fibre, which was then cut back to 5 m. Comparing the two transmitted spectra allowed the loss spectrum to be calculated [14] (Fig. 1(c)), giving a loss value of 0.062 dB/m at 2.8 μm. The fibre has a low loss (<0.08 dB/m) window between 2.5 and 3.5 μm, and the loss remains below 0.16 dB/m over the whole measurement range (2 to 4 μm). Next we measured the bending loss for different loop diameters $D$. For $D > 40$ cm, no bend-loss was observed in the wavelength range of 2.6 μm to 3.1 μm (the transmission window used in the experiment), confirming the excellent bending loss performance of hollow-core PCF [12, 31]. Using finite-element modelling (FEM), we calculated the dispersion to be −2.0 fs$^2$/mm at 2.8 μm (Fig. 1(c)).

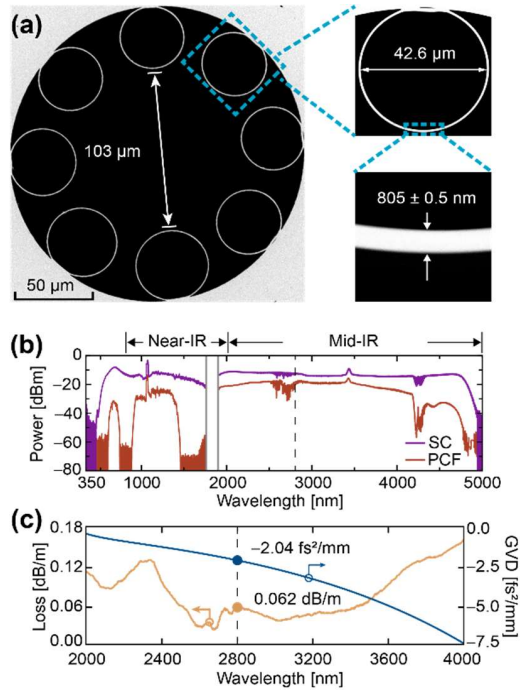

Fig. 1. The mid-IR hollow-core PCF. (a) Scanning electron micrograph of the fibre structure. (b) Spectrum (purple) of supercontinuum source and transmission spectrum (red) of hollow-core PCF. (c) Loss spectrum measured using the cut-back method (yellow, left axis), together with the calculated dispersion (blue, right axis).

## 3. Set-up and results of mid-IR ultrafast laser delivery

To investigate laser beam delivery (Fig. 2(a)), gas cells were attached to each end of a 5-m-long PCF sample, and connected to a vacuum pump. The light source, used in the laser-delivery experiment, was a mid-IR mode-locked fibre laser combined with a fibre amplifier [20], generating a pulse train with a maximum average power of ~1.1 W, a repetition rate of ~30 MHz, corresponding to a peak power of ~316 kW and a pulse duration of ~120 fs. The autocorrelation and optical spectrum of the pulses, measured using a mid-IR autocorrelator and a Fourier-transform-infrared (FTIR) spectrometer, show a pulse duration of ~117 fs FWHM and a 3-dB spectral bandwidth of ~127 nm (Figs. 2(c)&(e)). The time-bandwidth product of the laser pulse was calculated to be ~0.54, indicating some residual pulse chirp.

The laser light was launched into the hollow-core PCF using a coated $CaF_2$ plano-convex lens and a $CaF_2$ window, and the fibre was arranged in a loop of diameter ~75 cm. After optimisation, a launch efficiency of ~86% was obtained. As the fibre was gradually evacuated over 1 minute at a launched power of 780 mW, the transmission rose from ~45% to ~72%, resulting in ~563 mW transmitted power, and a near-diffraction-limited Gaussian profile (Fig. 2(b). The delivered power remained linearly proportional to the input power, rising to 779.8 mW for 1.1 W of input power, with almost constant transmission efficiency of ~70.7% (Fig. 2(b)). The autocorrelation and optical spectra of the transmitted pulses were measured at a launched power of 780 mW. In the diagnostics the free-space optical path was made as short as possible and both the autocorrelator and the FTIR spectrometer were purged with dry nitrogen gas to minimize atmospheric absorption (Fig. 2(a)).

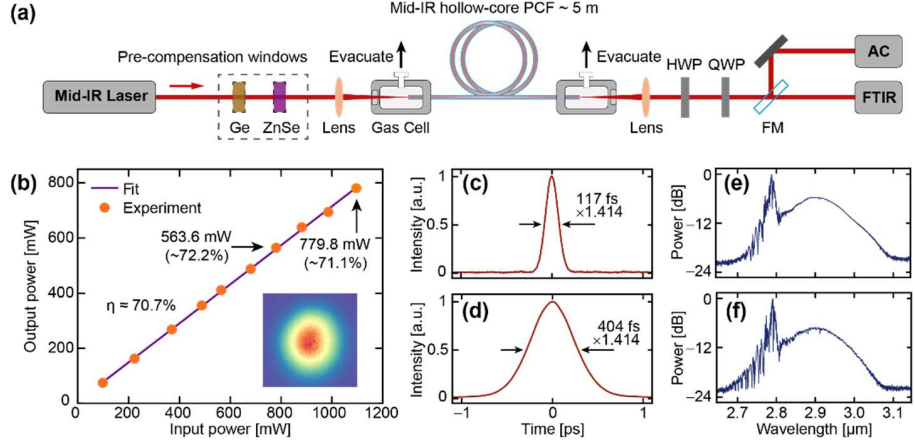

Fig. 2. The experimental setup and measurements. (a) Experimental set-up (see text for details). Lens, coated $CaF_2$ plano-convex lens; HWP, half-wave plate; QWP, quarter-wave plate; FM, flip mirror; OSA, optical spectrum analyzer; AC, autocorrelator. (b) Power transmitted through a 5 m length of hollow-core PCF as a function of input power. (c)&(d) Autocorrelation traces of the input and output pulses. (e)&(f) Spectra (normalized) of launched and transmitted pulses.

The autocorrelation trace and optical spectrum of the transmitted laser beam after the 5-m-long hollow-core PCF are shown in Figs. 2(d)&(f). The temporal pulse width was stretched to ~404 fs due to the waveguide dispersion of the PCF, while the output spectrum remained almost consistent with the input one, with the exception of some minor interference fringes observed (see Fig. 2(f)), which will be discussed in detail below. To analyze the temporal stretching, we employed in the simulation both the finite element modelling (FEM) and the analytical models to calculate the effective refractive index ($n_{eff}$) and the group velocity dispersion (GVD) curve of the fundamental optical mode in the PCF sample [12]. Using the capillary model, $n_{eff}$ can be expressed as:

$$n_{eff}(\omega) \approx 1 + \frac{1}{2}\left(-\left(\frac{2cu_{mn}}{\omega d}\right)^2 + \sum_q R_q(\omega)\right) \quad (1)$$

where $c$ is the speed of light in vacuum, $u_{mn}$ is the n[th] zero of the m[th] order Bessel function where $u_{01} \approx 2.405$, $\omega$ the light central frequency and $d$ the core diameter of the hollow waveguide. $R_q(\omega)$ is a Lorentzian-shaped response function used to describe the change of effective refractive index in the resonant region of the fibre. The simulated GVD curves of the hollow-core PCF (2 to 4 μm), based on FEM and capillary models, are shown in Fig. 3(a). The results exhibit good agreement, giving a GVD value of –2.0 $fs^2$/mm at 2.8 μm. Consequently, the 5-m-long hollow-core PCF has a total group delay dispersion (GDD) value of ~–10,000 $fs^2$, which can be pre-compensated using Ge, ZnSe and $CaF_2$ windows of varying thicknesses, as seen in Figs. 2(a) &3(b). Using Sellmeier equations [32-34], the material dispersion values at 2.8 μm can be calculated to be 1616 $fs^2$/mm (Ge), 180 $fs^2$/mm (ZnSe), –84 $fs^2$/mm ($CaF_2$) respectively.

We simulated the pulse propagation process in the hollow-core PCF, where the simulated pulse has a Gaussian shape with a full width at half maximum (FWHM) width of 135 fs (the increase of pulse width from 117 fs to 135 fs is mainly due to the coupling $CaF_2$ lens and the $CaF_2$ gas-cell windows). In the simulation, the input pulse spectrum supports a transform-limited pulse duration of ~96 fs and the calculated dispersion curve incorporates major higher-order dispersion terms. Figure 3(b) shows the simulated output pulse duration as a function of the additional GDD provided by the pre-compensation windows. Using combined windows with different materials and thicknesses, the pulse durations are also plotted as the blue dots in Fig. 3(b), exhibiting striking agreements with the simulated curve. The optimized pulse duration, obtained in the experiment, was ~98 fs, achieved by using a 30 mm ZnSe window and a 5 mm Ge window, which provide a total positive GDD value of $1.35\times10^4$ fs$^2$. These two windows effectively compensate for the residual chirp from the laser source and the negative GDD given by the hollow-core PCF, while introducing additional power loss of ~11%. Under this near-perfect-compensation condition, the autocorrelation traces of the output pulse obtained in the experiment (red curve) and in the simulation (blue curve) are shown in Fig. 5(c).

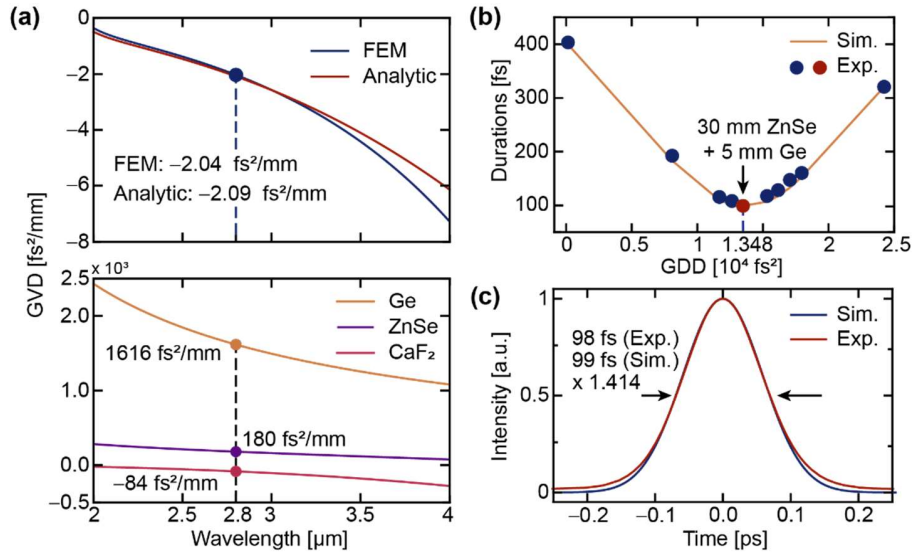

Fig. 3. The dispersion compensations and pulse durations measurements. (a) The GVD value of hollow-core PCF (top) calculated using FEM (blue) and capillary model (red). The GVD curves (bottom) for Ge (orange), ZnSe (purple), and CaF$_2$ (pink). (b) The shortest pulse duration (~98 fs) was achieved using 30-mm-thick ZnSe and 5-mm-thick Ge plates. (c) The experimentally (red) and calculated (blue) autocorrelation traces.

We measured the power, spectrum and pointing stabilities of both the mid-IR ultrafast laser and the output of the hollow-core PCF, as shown in Fig. 4. For assessing laser spectral stability, we recorded 2000 consecutive laser optical spectra using FTIR. Since each recording took roughly 2 second, the total measurement time for these 2000 sets of spectra was around 1 hours. It can be found in Figs. 4(a)&(b) that both of the laser beams before and after the fibre exhibited quite good spectral stabilities over the 1 hour recording period. To evaluate the power stability of the laser beams, we recorded the power of the laser direct output and the output power of the hollow-core PCF over 60 minutes using a mid-IR power meter, see Figs. 4(c)&(d). The laser power stability was be characterized using the coefficient of variation (CV) which is defined as the ratio of the standard deviation to the mean value. The calculated CV values for the input and the output laser beams were 0.45% and 0.47%, respectively, indicating excellent performance of the laser-delivery system on the power stability. For measuring the laser pointing-stability, both the input and output laser beams were focused onto a mid-IR camera using a CaF$_2$ lens with a focal length of 75 mm, with the centres of the laser beam profiles recorded over 1 hour. Shown in Figs. 4(e)&(f), the laser beam delivered by the hollow-core PCF exhibits better pointing stabilities ($\theta_x$ = 6.96 μrad and $\theta_y$ = 5.30 μrad, see Fig. 4(f)) than these of the input laser beam ($\theta_x$ = 14.33 μrad and $\theta_y$ =9.69 μrad, see Fig. 4(e)). This improved stability can be attributed to the enhanced waveguide and thermal characteristics of the hollow-core PCF [27].

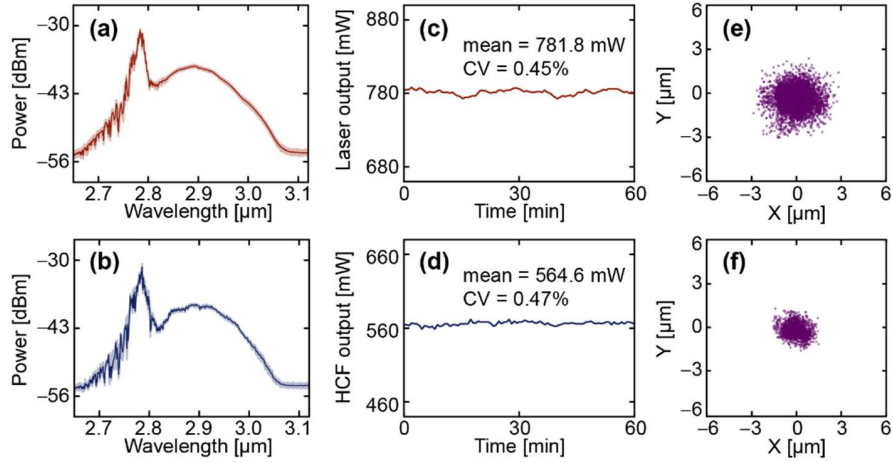

Fig. 4. Stabilities measurements. (a)&(b) The spectral stabilities of the hollow-core PCF input and output pulses, the solid line represents the average spectrum, and the shaded area describes the spectral fluctuation. (c)&(d) the power stability, (e)&(f) the pointing stabilities within 1 hour.

Minor interferometric fringes observed in the output spectrum arise from the interference between the fundamental ($LP_{01}$) and high-order ($LP_{11}$) optical modes supported by the hollow-core PCF, as shown in Fig. 5(a). To analyze this mode-beating phenomenon, we performed fast Fourier transform (FFT) on the output spectrum so as to obtain the autocorrelation function of the output pulse. As illustrated in Fig. 5(b), the FFT trace (purple line) correlates very well with the measured autocorrelation trace (red line), revealing the presence of a sub-pulse located at ~5.5 ps away from the main pulse. Using the FEM model, we simulated the group velocities of the $LP_{01}$ and $LP_{11}$ modes in the hollow-core PCF, and the simulation results are illustrated in Fig. 5(c). The group velocity difference between the two modes is ~1.04 ps/m, giving rise to 5.2 ps group delay over 5-m-long propagation in hollow-core PCF, which agrees well with the experimental value of 5.5 ps. From the measured autocorrelation trace, we can estimate that the main pulse (in $LP_{01}$ mode) contains >95% of the total energy, indicating the excellent single-mode performance of the hollow-core PCF [31].

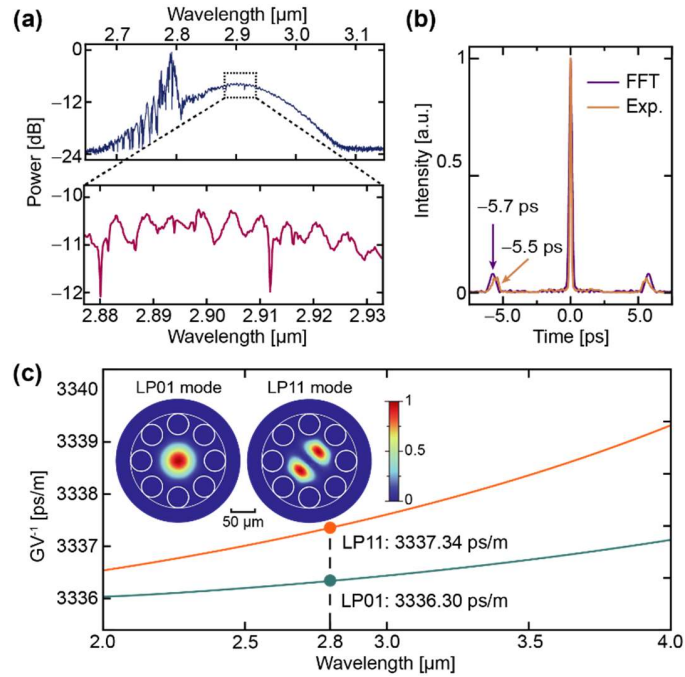

Fig. 5. Mode-purity characterization. (a) The spectral fringes of hollow-core PCF outputs. (b) The Fourier transforms of the fibre output spectrum (purple) and the experimental measured autocorrelation trace (orange). (c) The group delay parameter curves of the $LP_{01}$ mode (green) and $LP_{11}$ mode (orange). The inset shows the simulated $LP_{01}$ and $LP_{11}$ modes of the hollow-core PCF.

## 4. Comparison with results of free-space or air-filled-PCF transmission

The laser spectrum contains both the v1 and v3 vibrational bands of water vapor, which exhibit strong absorption lines within 2.7 – 2.9 μm regime, as shown in Fig. 6(a). Such water absorption would cause substantial loss of the transmitted power, additionally it would also result in significant distortions of the laser temporal and spatial profiles [19]. For comparison, in the experiment, mid-infrared ultrafast light was propagated over a 5-m-long free-space path under a relative humidity of approximately 70% at 24°C (corresponding to an absolute humidity of 15.23 g/m³). With ~780 mW of laser power launched into the 5-m free-space path, the measured transmitted power was ~390 mW, yielding an efficiency of ~50%, accompanied by a notably distorted spectrum featuring dense absorption lines.

In the case of the hollow-core PCF, vacuum pumping is essential to eliminate both the absorption and nonlinearity caused by the ambient air. As illustrated in Fig. 6(c), the hollow-core PCF operating under conditions open to the laboratory environment is unable to avoid water absorption. Under these conditions, at the same laser power of ~780 mW, the transmitted power was measured to be merely ~350 mW, giving a relatively-low transmission efficiency of ~45%. The absorption lines can be effectively eliminated by evacuating the PCF to ~10 mbar within 1 minute. During evacuation the transmitted laser power increased quickly to ~565 mW, see Fig. 6(d) for the transmitted laser spectrum after evacuation.

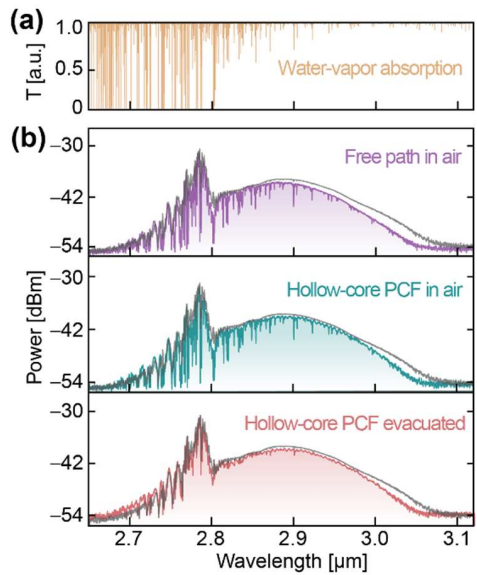

Fig. 6. The spectra measurements by water-vapor influence. (a) Normalized absorption spectrum of water vapor. (b) Spectra of direct laser output spectrum (gray) and the transmitted spectrum in a ~5-meter free path in air (purple), the hollow-core PCF output within air (green), and the hollow-core PCF output within vacuum (red).

## 5. Discussion and conclusions

In this work, we demonstrated the high-quality delivery of near-Wat-level 2.8 μm ultrafast laser pulses through an evacuated hollow-core PCF sample, achieving remarkably-high spectral, temporal and spatial fidelities. The fused-silica material of the hollow-core PCF improves the mechanical and chemical robustness of the fibre compared to its counterparts made from mid-IR materials, while its low-loss performance of the fibre gives rise to relatively-high transmission efficiency of >70%. Additionally, a straightforward dispersion pre-compensation design using ZnSe and Ge windows ensures excellent temporal performance, with pulse widths at the 100-fs level. The use of vacuum pumping enables near-distortion-free transmission of the broadband mid-IR spectrum, even across the water-absorption bands. The delivered laser beam features outstanding spectral and power stabilities, with, surprisingly, improved pointing stability thanks to the waveguide characteristics of the hollow-core PCF.

The remaining issue lies in that a tiny portion (~4%) of transmitted energy was coupled into the high-order optical mode ($LP_{11}$) of the fibre, which could be suppressed through the implementation of new hollow-core PCF designs [31]. Moreover, the current set-up is certainly capable of supporting longer transmission lengths, mainly due to the advancements in hollow-core PCF fabrication techniques that enhance fibre uniformity and reduce loss [12, 15]. The results presented here pave the way for advanced applications of broadband mid-IR ultrafast light sources in spectroscopy, imaging, bio-photonics and remote sensing, all of which would be benefited from the flexible fibre transmission of mid-IR laser beams with high spectral, temporal and spatial fidelities.

**Funding.** Strategic Priority Research Program of the Chinese Academy of Science (XDB0650000); Shanghai Science and Technology Innovation Action Plan (21ZR1482700); Shanghai Science and Technology Plan Project Funding (23JC1410100); National High-level Talent Youth Project; Fuyang High-level Talent Group Project.

**Disclosures.** The authors declare no conflicts of interest.

**Data availability.** Data underlying the results presented in this paper are not publicly available at this time but may be obtained from the authors upon reasonable request.

## References

1. D. A. Long, M. J. Cich, C. Mathurin, *et al*., "Nanosecond time-resolved dual-comb absorption spectroscopy," Nat. Photonics **18**, 127-131 (2023).
2. A. V. Muraviev, V. O. Smolski, Z. E. Loparo, *et al*., "Massively parallel sensing of trace molecules and their isotopologues with broadband subharmonic mid-infrared frequency combs," Nat. Photonics **12**, 209-214 (2018).
3. A. Schliesser, N. Picqué, and T. W. Hänsch, "Mid-infrared frequency combs," Nat. Photonics **6**, 440-449 (2012).
4. A. J. Lind, A. Kowligy, H. Timmers, *et al*., "Mid-Infrared Frequency Comb Generation and Spectroscopy with Few-Cycle Pulses and χ(2) Nonlinear Optics," Phys. Rev. Lett **124**, 133904 (2020).
5. D. Bubb, J. Horwitz, R. McGill, *et al*., "Resonant infrared pulsed-laser deposition of a sorbent chemoselective polymer," Appl. Phys. Lett **79**, 2847-2849 (2001).
6. S. M. O'Malley, J. Schoeffling, R. Jimenez, *et al*., "The influence of wavelength, temporal sequencing, and pulse duration on resonant infrared matrix-assisted laser processing of polymer films," Appl. Phys. A **117**, 1343-1351 (2014).
7. A. Urich, R. R. Maier, F. Yu, *et al*., "Flexible delivery of Er:YAG radiation at 2.94 μm with negative curvature silica glass fibers: a new solution for minimally invasive surgical procedures," Biomed. Opt. Express **4**, 193-205 (2013).
8. P. Krogen, H. Suchowski, H. Liang, *et al*., "Generation and multi-octave shaping of mid-infrared intense single-cycle pulses," Nat. Photonics **11**, 222-226 (2017).
9. M. Fernandez-Vallejo, and M. Lopez-Amo, "Optical fiber networks for remote fiber optic sensors," Sensors (Basel) **12**, 3929-3951 (2012).
10. G. B. Rieker, F. R. Giorgetta, W. C. Swann, *et al*., "Frequency-comb-based remote sensing of greenhouse gases over kilometer air paths," Optica **1**, 290-298 (2014).
11. K. Zheng, S. Jiang, F. Chen, *et al*., "Mid-infrared all-optical modulators based on an acetylene-filled hollow-core fiber," Light Adv. Manuf **3**, 712-719 (2022).
12. E. Numkam Fokoua, S. Abokhamis Mousavi, G. T. Jasion, *et al*., "Loss in hollow-core optical fibers: mechanisms, scaling rules, and limits," Adv. Opt. Photonics **15**, 1-85 (2023).
13. H. C. H. Mulvad, S. Abokhamis Mousavi, V. Zuba, *et al*., "Kilowatt-average-power single-mode laser light transmission over kilometre-scale hollow-core fibre," Nat. Photonics **16**, 448-453 (2022).
14. Q. Fu, Y. Wu, I. A. Davidson, *et al*., "Hundred-meter-scale, kilowatt peak-power, near-diffraction-limited, mid-infrared pulse delivery via the low-loss hollow-core fiber," Opt. Lett. **47**, 5301-5304 (2022).
15. P. Uebel, M. C. Günendi, M. H. Frosz, *et al*., "Broadband robustly single-mode hollow-core PCF by resonant filtering of higher-order modes," Opt. Lett. **41**, 1961-1964 (2016).
16. G. Tao, H. Ebendorff-Heidepriem, A. M. Stolyarov, *et al*., "Infrared fibers," Adv. Opt. Photonics **7**, 379-458 (2015).
17. J. Ballato, T. Hawkins, P. Foy, *et al*., "Silica-clad crystalline germanium core optical fibers," Opt. Lett. **36**, 687-688 (2011).
18. I. Vitoria, C. R. Zamarreño, A. Ozcariz, *et al*., "Beyond near-infrared lossy mode resonances with fluoride glass optical fiber," Opt. Lett. **46**, 2892-2895 (2021).
19. M. Gebhardt, C. Gaida, F. Stutzki, *et al*., "Impact of atmospheric molecular absorption on the temporal and spatial evolution of ultra-short optical pulses," Opt. Express **23**, 13776-13787 (2015).
20. J. Huang, M. Pang, X. Jiang, *et al*., "Sub-two-cycle octave-spanning mid-infrared fiber laser," Optica **7**, 574-579 (2020).
21. F. Désévédavy, G. Renversez, J. Troles, *et al*., "Chalcogenide glass hollow core photonic crystal fibers," Opt. Mater **32**, 1532-1539 (2010).
22. A. Ventura, J. G. Hayashi, J. Cimek, *et al*., "Extruded tellurite antiresonant hollow core fiber for Mid-IR operation," Opt. Express **28**, 16542-16553 (2020).
23. H. Zhang, Y. Chang, Y. Xu, *et al*., "Design and fabrication of a chalcogenide hollow-core anti-resonant fiber for mid-infrared applications," Opt. Express **31**, 7659-7670 (2023).
24. A. N. Kolyadin, A. F. Kosolapov, A. D. Pryamikov, *et al*., "Light transmission in negative curvature hollow core fiber in extremely high material loss region," Opt. Express **21**, 9514-9519 (2013).
25. Q. Fu, I. A. Davidson, S. M. A. Mousavi, *et al*., "Hollow-Core Fiber: Breaking the Nonlinearity Limits of Silica Fiber in Long-Distance Green Laser Pulse Delivery," Laser. Photonics. Rev **18**, 2201027 (2024).
26. M. A. Cooper, J. Wahlen, S. Yerolatsitis, *et al*., "2.2 kW single-mode narrow-linewidth laser delivery through a hollow-core fiber," Optica **10**, 1253-1259 (2023).
27. C. Yan, H. Li, Z. Huang, *et al*., "Highly stable, flexible delivery of microjoule-level ultrafast pulses in vacuumized anti-resonant hollow-core fibers for active synchronization," Opt. Lett. **48**, 1838-1841 (2023).
28. W. Hongya, A. Jianzhou, M. Zelin, *et al*., "Finding the superior mode basis for mode-division multiplexing: a comparison of spatial modes in air-core fiber," Adv. Photonics **5**, 056003 (2023).
29. F. Yu, P. Song, D. Wu, *et al*., "Attenuation limit of silica-based hollow-core fiber at mid-IR wavelengths," APL Photonics **4**, 080803 (2019).
30. L. Cao, S. Gao, Z. Peng, *et al*., "High peak power 2.8 μm Raman laser in a methane-filled negative-curvature fiber," Opt. Express **26**, 5609-5615 (2018).

31. Z. Luo, J. Huang, Y. Zheng, *et al*., "Ultrahigh Transverse Mode Purity by Enhanced Modal Filtering in Double-Clad Single-Ring Hollow-Core Photonic Crystal Fiber," Laser. Photonics. Rev **18**, 2301111 (2024).
32. T. Amotchkina, M. Trubetskov, D. Hahner, *et al*., "Characterization of e-beam evaporated Ge, YbF(3), ZnS, and LaF(3) thin films for laser-oriented coatings," Appl. Opt **59**, A40-A47 (2020).
33. H. H. Li, "Refractive index of alkaline earth halides and its wavelength and temperature derivatives," J. Phys. Chem. Ref. Data **9**, 161-290 (1980).
34. J. Connolly, B. diBenedetto, and R. Donadio, "Specifications Of Raytran Material," Proc. SPIE **0181**, 141-144 (1979).